\documentclass[twocolumn,showpacs,preprintnumbers,amsmath,amssymb]{revtex4}

\usepackage{graphicx}% Include figure files
\usepackage{bm}% bold math

\begin{document}

\title{Investigation of a ring single molecular magnet Mn$_6$R$_6$ in
megagauss fields}

\author{ V.V. Kostyuchenko$^1$, I.M. Markevtsev$^2$, A.V.
Philippov$^2$,
V.V.Platonov$^2$, V.D. Selemir$^2$, O.M. Tatsenko$^2$, A.K.
Zvezdin$^3$, A. Caneschi$^4$}
\affiliation{
$^1$ Institute of Microelectronics and Informatics RAS, Yaroslavl, Russia\\
$^2$ VNIIEF, Sarov, Russia\\
$^3$General Physics Institute RAS, Moscow, Russia\\
$^4$ Dipartimento di Chimica e UdR INSTM of Firenze, Italy.}

\date{\today}

\begin{abstract}
The dependence of the magnetic susceptibility on the magnetic
field is investigated for the single molecular magnet
[Mn(hfac)$_2$NITPh]$_6$. The spikes of the susceptibility
detected in a magnetic field ranging from 90 T to 285 T are
interpreted as a manifestation of magnetic quantum jumps under
spins reorientation from ferrimagnetic to ferromagnetic
structure. The characteristic feature of the single molecular
magnet Mn$_6$ R$_6$ is a deficiency of pair Heisenberg exchange
interactions for the description of its magnetic properties in
high magnetic fields. The comparison of the experimental data
with the results of theoretical calculations allows us to prove
the existence of strong three-spin interaction in this molecular
cluster and to determine the values of exchange constants. For
the calculation of the ground state spin structure the modified
Lanczos method is used.
\end{abstract}


\maketitle

\section*{Introduction}
Molecular magnetism is one of the most developing fields in the
physics of magnetism. Magnetic molecular nanoclusters containing
transition metal ions are an important component in this area of
investigations \cite{Chudn*, phtd:12-35, ccmp:17-39, epl:27-159,
Kahn, jmmm:200-167, ufn:166-439, ftt:42-1068, ftt:43-177,
jetp:109-2115}. On the one hand, these materials allow the
investigation of macroscopic quantum phenomena \cite{prb:30-1208,
sci:258-414, jap:73-6709, jap:73-6703, sptp:69-80}, on the other
hand, molecular magnetic clusters are promising components for
the design of new magnetic materials. Furthermore, the study of
these materials is important due to the possibility of their
utilization in magnetocaloric and magnetooptic devices
\cite{prl:68-3749, jmmm:85-219, nat:268-437}, or in quantum
computers (see e.g. \cite{rmp:68-733, rpp:61-117, ufn:169-507}
and references therein).

In the present work the interesting cluster
[Mn(hfac)$_2$NITPh]$_6$ (see \cite{jacs:110-2795}) is investigated.
For brevity sake we denote it as Mn$_6$R$_6$ below. The single molecular
magnet
Mn$_6$R$_6$ being synthesized in 1988 forms dark green hexagonal
crystals. According to \cite{jacs:110-2795} the crystals of
Mn$_6$R$_6$ have rhombohedral group of symmetry $R\bar 3$ (the cell
parameters are $a=b=c=21.21$ \AA,
$\alpha=\beta=\gamma=116.77^\circ$). The molecular structure of
Mn$_6$R$_6$ is shown in Figure 1. The spin structure (see Figure 2)
of the molecular cluster consists of a ring of six ions $Mn$ (spin $S_{Mn}$=5/2)
alternating with radicals ($S_{R}$=1/2). In the ground state the
total spin of the molecular cluster has a value of $S$=12 which results
from an antiferromagnetic exchange interaction between $Mn$ ions and
radicals. The spikes detected in the dependence of magnetic
susceptibility on the external magnetic field can be interpreted
as a stepwise changing of the total spin in the ground state. The
increase of the total spin necessarily gives rise to an increase of
the antiferromagnetic exchange energy between $Mn$ ions and $R$ radical units. So
this transition occurs only if the Zeeman energy of the spins compensate
the increase of the exchange energy.

The study of the transitions induced by external magnetic field is a
direct method to investigate the exchange interaction. As long as the
interaction between spins of separate molecules in single
molecular magnets is small these materials behave like an ensemble
of non-interacting microscopic objects.

In the present work the experimental investigations of magnetic
susceptibility of single molecular magnet Mn$_6$R$_6$ are made
along with numerical calculations of the energy of the molecular
cluster in the ground state. The obtained experimental data are
compared with the results of the theoretical calculations and it
allows to prove that an acceptable agreement between the
experimental data and the theoretical ones can only be obtained
if multi-spin interactions are taken into account. Taking into
consideration the three-spin interaction allows us to obtain a
rather good agreement of experimental and theoretical data
(discrepancy is less than 10\%).

\section{Experiment}
To investigate the single molecular magnet Mn$_6$R$_6$ in
magnetic fields up to 400 T the explosive generator MC-1 was used.
The generator works on the basis of the compression principle of
the conducting shell containing magnetic field inside. In our
experiments a single cascade generator with initial cascade diameter
139 mm was used. The initial magnetic field in the experiments was 16 T. The
detected value of magnetic field in maximum comes up to 503 T.
During the process of generation the magnetic field registration
was performed by several detectors representing the one loop coils
with various diameters. The use of coils with various diameters
allows the more accurate detection of field derivatives.
The MC-1 generator is shown in Figure 3.

The compensation method was used as experimental procedure for measurements of changing magnetization. The
essence of this technique is that of using the coil in which half turns are wound on tube with
right-hand thread and other half turns are wound on tube with
left-hand thread. If the coils are placed in changing magnetic field the
EMF of opposite polarity and same value are induced in the coils
giving rise to the total EMF $\sim$ 0. If a sample is placed in one
arm of measuring coil the total EMF is different from zero. The
signal is proportional to the changing of the magnetization of the sample.
In experiments coils with inner diameter 1.6 mm and a number of turns equal to
18 are used. The schmeme of such coil is given in Fig.ure 4.

The sample was placed inside the compensation detector and was
immediately sealed hermetically by glue. As long as the
experiment was made at the temperature of liquid helium the
compensation detector along with the sample was placed in the
cryostat. The flowing cryostat was used allowing to maintain a
sample temperature of 4.2 K.

\section{Hamiltonian and results of numerical calculations}

The spin structure of the single molecular magnet is made up by a ring formed by
six $Mn$ ions with spins $S_{Mn}=5/2$ alternating with radicals
(spin $S_{R}=1/2$). According to the experimental data obtained in
\cite{jacs:110-2795} of the magnetic susceptibility,
the single molecular magnet $Mn_6 R_6$ has a total spin of
$S_\Sigma=12$ in the ground state at $H=0$. This can be explained by the
existence of antiferromagnetic exchange interaction between $Mn$
ions and $R$ radicals.
\begin{equation} \label{1}
{\cal H}_A=J\sum_{i=1}^{12}\vec S_i\vec S_{i+1},
\end{equation}
\noindent In equation (\ref{1}) and below it is assumed that even
indexes correspond to the $Mn$ ions and odd indexes to the
radicals. The coupling scheme of the spin structure in the
absence of an external magnetic field is given in Figure 2.
According to the data present in literature \cite{jacs:110-2795}
the coupling constant $J \sim 10^2$ cm$^{-1}$.

The change of the cluster spin structure in the external magnetic
field is caused by the change of its Zeeman energy
\begin{equation} \label{2}
{\cal H}_Z=g\mu_B H\sum_{i=1}^{12}S_i^z,
\end{equation}

\noindent  $g$ -- g-factor, $\mu_B$ -- Bohr magneton. When a stepwise
change of the spin structure of the cluster occurs, the increase of
the exchange interaction energy is compensated by the decrease of the
Zeeman energy. Given values of magnetic field correspond to the
spikes found in the behavior of magnetic susceptibility in dependence on magnetic field variation
$\chi\left( H \right)$.

The experimental dependence of $\chi\left( H \right)$ is shown in
Figure 5. From this Figure is possible to appreciate that spikes
of susceptibility are grouped in a range of fields from 90 T up
to $\sim$ 285 T. It is apparent that these spikes correspond to
the jumps of magnetization during the transition from
ferrimagnetic phases to the ferromagnetic ones. Using different
words they can be interpreted as transitions of a "low spin --
high spin" type, induced by the external magnetic field level
intersection in the ground state. It seems natural an attempt to
describe these transitions on the base of the simplest model ---
the Heisenberg model
\begin{equation} \label{3}
{\cal H}_2={\cal H}_A+{\cal H}_Z.
\end{equation}

The performed calculations show that in the framework of the
Heisenberg model it is impossible to obtain acceptable agreement
between theoretical calculations and the experimental data. This
is apparent even from simple estimation of the critical fields of
transition from ferrimagnetic state to the ferromagnetic one.
According to the familiar formula (see e.g. \cite{Zvezd}) the
critical fields for two sublattice ferrimagnet are given by
\begin{equation} \label{4}
H_{C_{1,2}}=J\left(M_1 \pm M_2 \right),
\end{equation}

\noindent here $J$ -- exchange constant, $M_i$ -- sublattice
magnetizations. As follows from eq. (\ref{4}) the ratio between the width
of the reorientation range $\Delta H$ and the value of the field in its
center equals
\begin{equation} \label{5}
\frac{\Delta H}{H_0}=\frac{2 M_2}{M_1}.
\end{equation}
For Mn$_6$R$_6$ (sublattice magnetizations are $\mu_B$ and
5$\mu_B$) this value is 0.4 whereas its experimental value is
$\sim$1. Although the process of spin reorientation in
Mn$_6$R$_6$ is not a continuous but a quantum (discrete) one, the
relation presented above can be used for qualitative estimation.
The strong difference between the experimental data and the
theoretical approximation based upon the Heisenberg model
indicates that this model is not sufficient for the
interpretation of the experiment. It is obvious that taking into
account anisotropy energy as well does not help to solve the
problem (another reason is that the Mn$^{2+}$ ions are
$S$--ions).

The characteristic feature of the experimental dependence
$\chi\left( H \right)$ is a strong change of the distances between
adjacent spikes. The change of the interval between adjacent spikes
can be estimated on the base of spin-wave approximation (see e.g.
\cite{Barya*}). For the chain of identical spins the ratio between
the change of the interval and its value is of the order of $1/NS$
\cite{Barya*} and for chains of 12 spins can reach several
percents. We performed the analogous calculation for
a ferrimagnetic chain: it gives rise to more difficult analytical
relations but the order of the value does not exceed 10 percent as
well. It is apparent that taking into account direct exchange
interaction between $Mn$ ions or between radicals also does not
give rise to essential change of the intervals between adjacent
spikes.

The results of the analytical approximation were verified by
numerical calculations. Unlike to the another molecular clusters
investigated by the perturbation theory \cite{ftt:42-1068,
ftt:43-177, jetp:109-2115} in the present case the perturbation
theory is not applicable. The modified Lanczos method (see e.g.
\cite{prb:34-1677, rmp:66-763}) was used for the calculation of
Mn$_6$R$_6$ spin structure and energy in the ground state. The
choice of algorithm is explained by the necessity of calculations
of the ground state energy and structure in the spin subspace of
large dimensions (up to 5330), for which a large amount of
calculations is needed. Modified Lanczos method demands for
energy calculation of the ground state several times less matrix
multiplication than another methods. Again, its reliability (it
does not arise accumulation of errors during process of
calculations) allows to create the compact program code to find
the solution of this problem with high efficiency.

The results of the numerical calculations are in complete
agreement with the analytical estimations. On the base of the
Hamiltonian (\ref{3}), as in the case of taking into account the
exchange interactions between nearest $Mn-Mn$ ions and $R-R$
radicals does not succeed to obtain acceptable agreement of
experimental data and the results of numerical calculations. The
change of the interval between adjacent spikes does not exceed
10-15 \% at any reasonable choice of exchange constants. Taking
into account the anisotropy of the exchange Heisenberg
interaction also does not allow an explanation for the strong
change of the intervals between adjacent spikes. Thus, for the
interpretation of the experimental data it is necessary to take
into consideration non-Heisenberg exchange interactions
(biquadratic, three-spin etc.). As far as the spin of the radical
is equal to 1/2 then using well known property of Pauli matrixes
\begin{equation}
\left(\vec \sigma \vec A\right) \left(\vec \sigma \vec B\right) =
\left(\vec A\vec B\right) +i\vec \sigma \left[ \vec A\times \vec
B\right],
\end{equation}

\noindent it is easy to show that taking into account biquadratic
exchange interaction of type $\left(\vec S_i\vec
S_{i+1}\right)^2$ and three-spin interaction of type $\left(\vec
S_{2i}\vec S_{2i+1}\right)\left(\vec S_{2i+1}\vec
S_{2i+2}\right)$ it gives rise only to renormalization of the
constants of Heisenberg exchange interaction. This means that
among all non-Heisenberg two-spin and three-spin interactions
only taking into account the three-spin interaction of the type
\begin{multline} \label{6}
%\begin{array}{c}
{\cal H}_3=J_3\sum_{j=1}^6\left[ \left( \vec S_{2j-1}\vec
S_{2j}\right) \left( \vec S_{2j}\vec S_{2j+1}\right) +
\right.\\
\left.
\left(\vec S_{2j}\vec S_{2j+1}\right) \left( \vec S_{2j-1}\vec
S_{2j}\right) \right],
%\end{array}
\end{multline}

\noindent may be essential. As shown below taking into
consideration the three-spin interaction of the type (\ref{6})
really allows to explain the experimental data.

For the calculation of the positions of the spikes we used the Hamiltonian
\begin{equation} \label{7}
{\cal H}={\cal H}_A+{\cal H}_Z+{\cal H}_3.
\end{equation}

\noindent The positions of the spikes are calculated from the condition of
equality for the minimal energies of states with various total spins at
various values of $J_3/J$ ratio. Next, the results of the calculations
were compared with experimental data. Every spike on the
dependence of magnetic susceptibility on external magnetic field
corresponds to an increase of the total spin of the molecular cluster
$Mn_6 R_6$ by 1. The change of the total spin from $S_\Sigma =12$ to
$S_\Sigma =18$ must then correspond to the 6 spikes present in the curve
$\chi\left( H \right)$. As the instrumental resolution is
less in the fields higher than 250 T, the three last spikes in this
curve are merged into one broad spike. It makes the comparison of
experimental and theoretical data more difficult.

In the limiting cases the position of fourth spike in the
experimental curve $\chi\left( H \right)$ can be associated with
the transition $S_\Sigma =16 \to S_\Sigma =17$ or with the
transition $S_\Sigma =15 \to S_\Sigma =16$. In the first case the
best agreement with experimental data is achieved at the
following values of exchange constants: $J=27 cm^{-1}$ and
$J_3/J=0.05$. The discrepancy between experimental and
theoretical data is 24 \%. In the second case the exchange
constants equal $J=40 cm^{-1}$ and $J_3/J=0.14$ and the
dispersion is less than 10 \% (see Table 1). The obtained value
of the Heisenberg constant is approximately two times smaller
than the estimated value in \cite{jacs:110-2795}. It can be
explained by the strong delocalization of the corresponding spin
density of the radicals and, as a consequence, by the strong
three-spin interaction. The strong delocalization of the spin
density on part of the skeleton radical molecule was
experimentally showed for this radiacal and similar derivative in
\cite{jacs:116-2019, chem:5-3616}. Thus it can be concluded that
the position of the fourth spike in the experimental curve
$\chi\left( H \right)$ is close to the value of the magnetic
field corresponding to the transition $S_\Sigma =15 \to S_\Sigma
=16$. It should be pointed out that in this case the value of the
three-spin interaction is rather large.

\section*{Conclusions}
In this work experimental and theoretical investigations of the
magnetic susceptibility of the single molecular magnet
[Mn(hfac)$_2$NITPh]$_6$ are presented. It is shown that
acceptable agreement between experimental data and results of
theoretical calculations can be obtained only by taking into
account three-spin interaction. In this case the value of the
three-spin interaction is rather large and the ratio of
three-spin exchange constant and Heisenberg exchange constant may
be as much as 0.14.

\section*{Acknowledgements}
This work supported by grants MNTP (project N 97-1071), RFBR
(99-02/17830) and INTAS (99-01839).

\begin{table}[h]
\vskip 1 cm \centering
\begin{tabular}{|c|c|c|c|c|c|c|}
 \hline\hline
 $H_{exp}$, T & 92 & 134 & 198 & 284 & -- & -- \\
 \hline
 $H_{theor}$, T & 104 & 126 & 178 & 303 & 322 & 327 \\
\hline \hline
\end{tabular}
\caption{Experimental and theoretical values of magnetic
susceptibility spikes for $J=39.8 cm^{-1}$ and $J_3/J=0.14$.}
\end{table}

\begin{figure*}[h]
\includegraphics[width=0.8\textwidth]{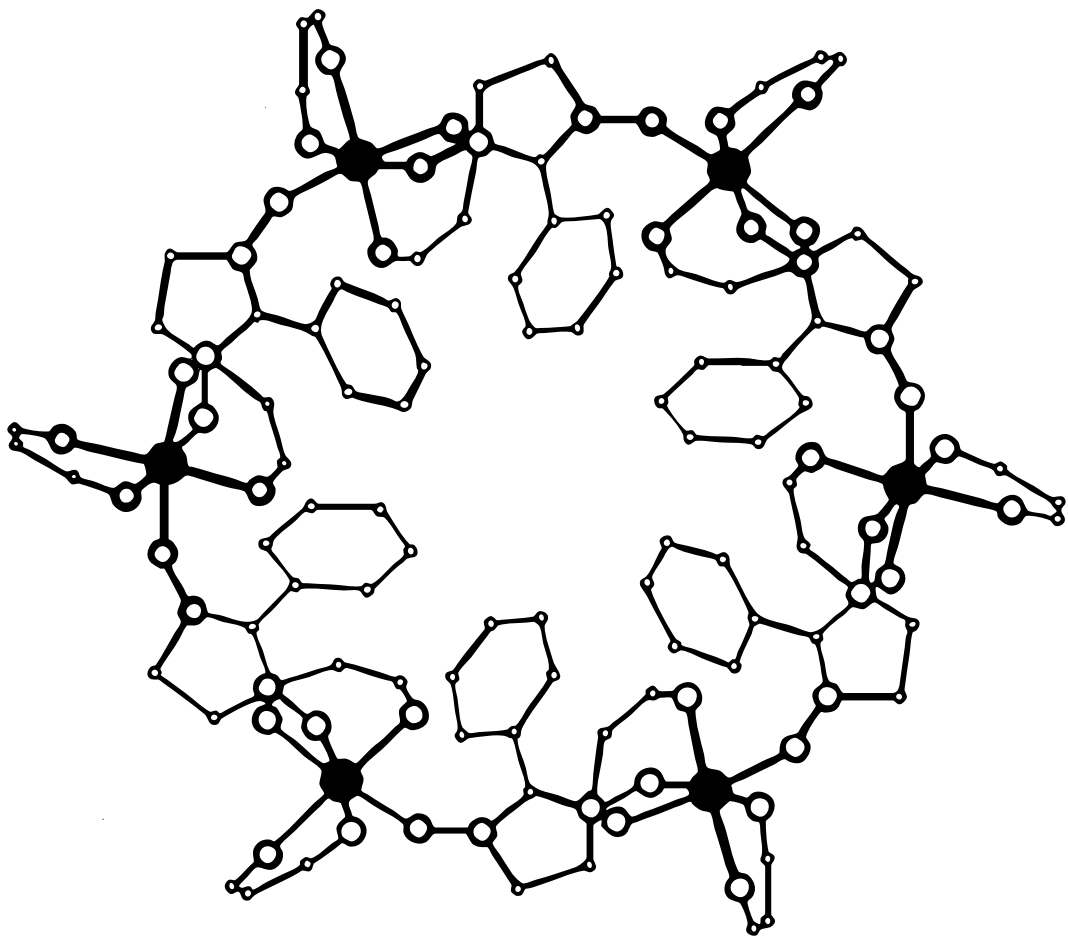}
\caption{The molecular structure of single molecular magnet
Mn$_6$R$_6$.}
\end{figure*}

\begin{figure*}[h]
\includegraphics[width=0.8\textwidth]{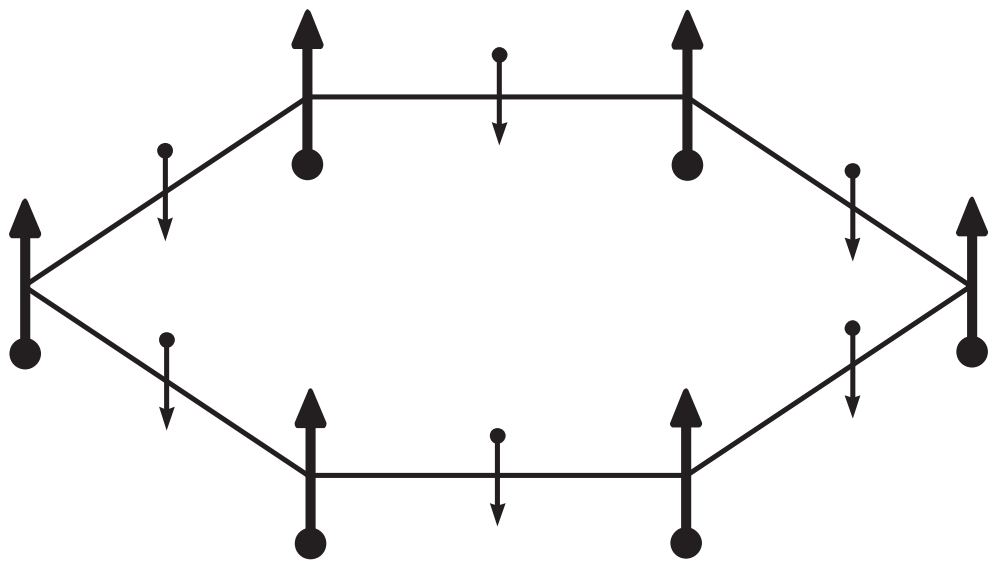}
\caption{The scheme of spin structure in Mn$_6$R$_6$.}
\end{figure*}

\begin{figure*}[h]
\includegraphics[width=0.8\textwidth]{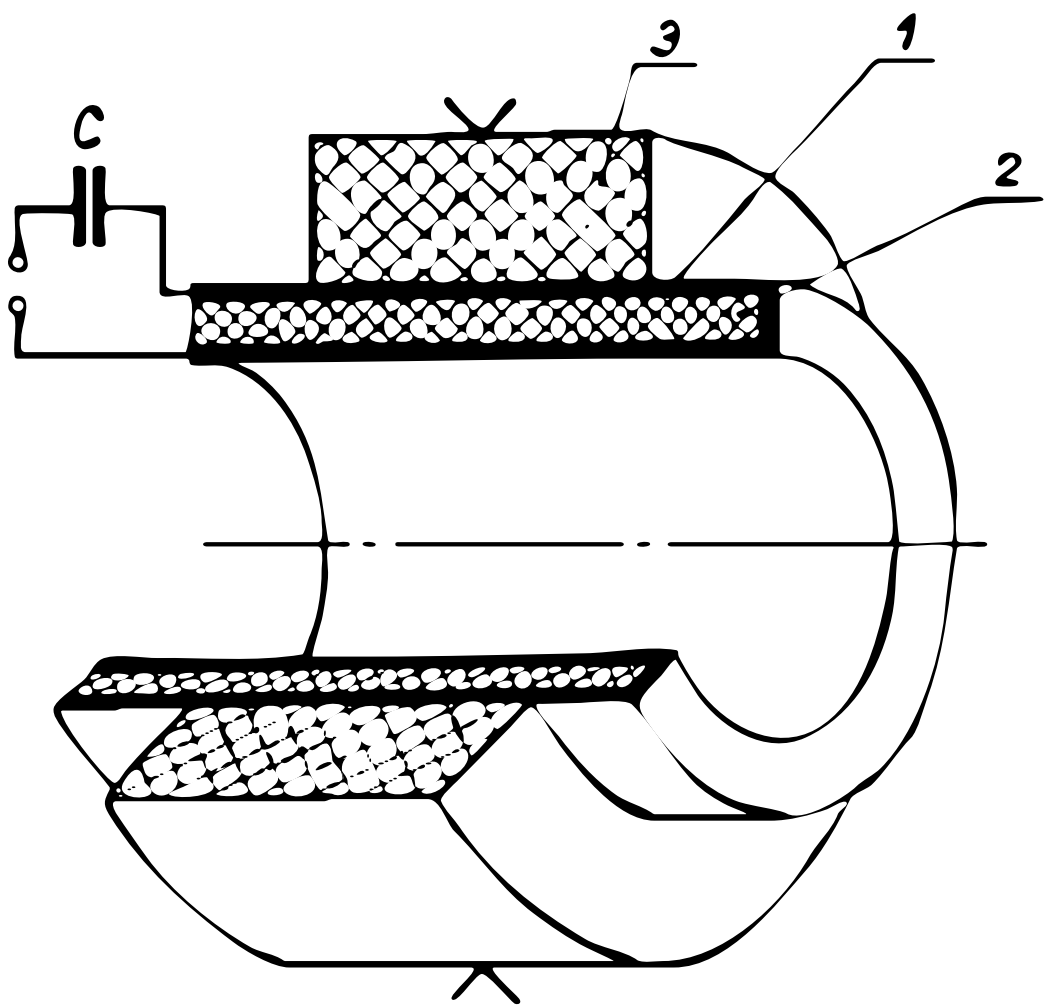}
\caption{The scheme of MC-1 generator.}
\end{figure*}

\begin{figure*}[h]
\includegraphics[width=0.8\textwidth]{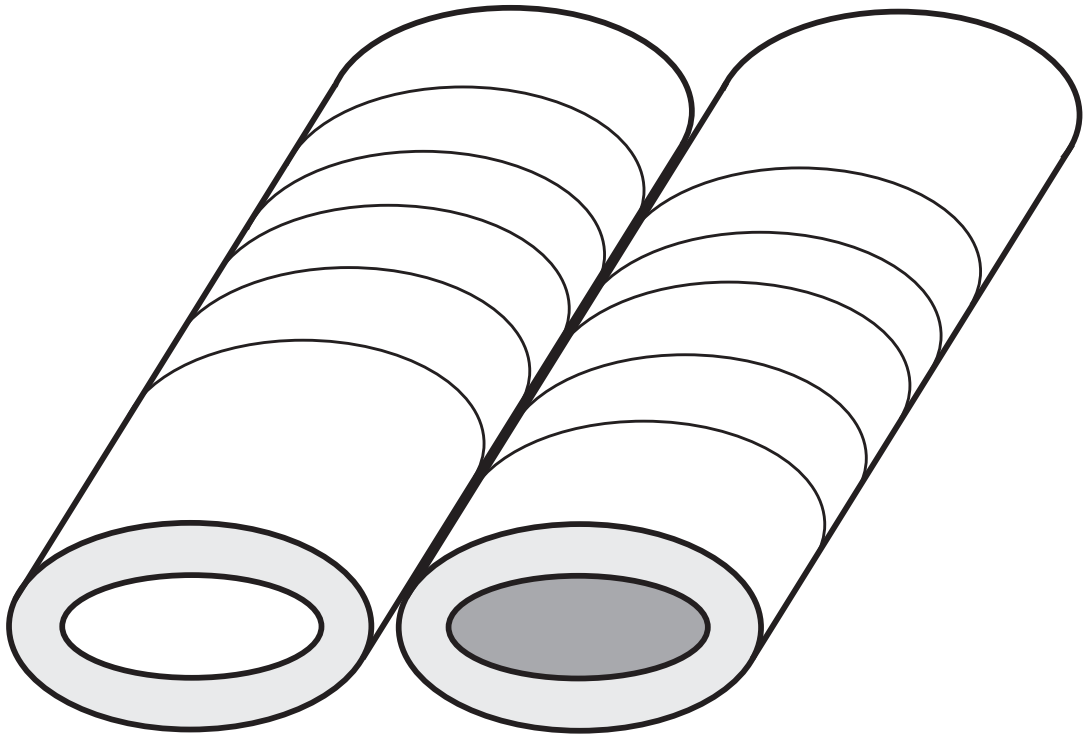}
\caption{The view of measuring coil.}
\end{figure*}

\begin{figure*}[h]
\includegraphics[width=0.8\textwidth]{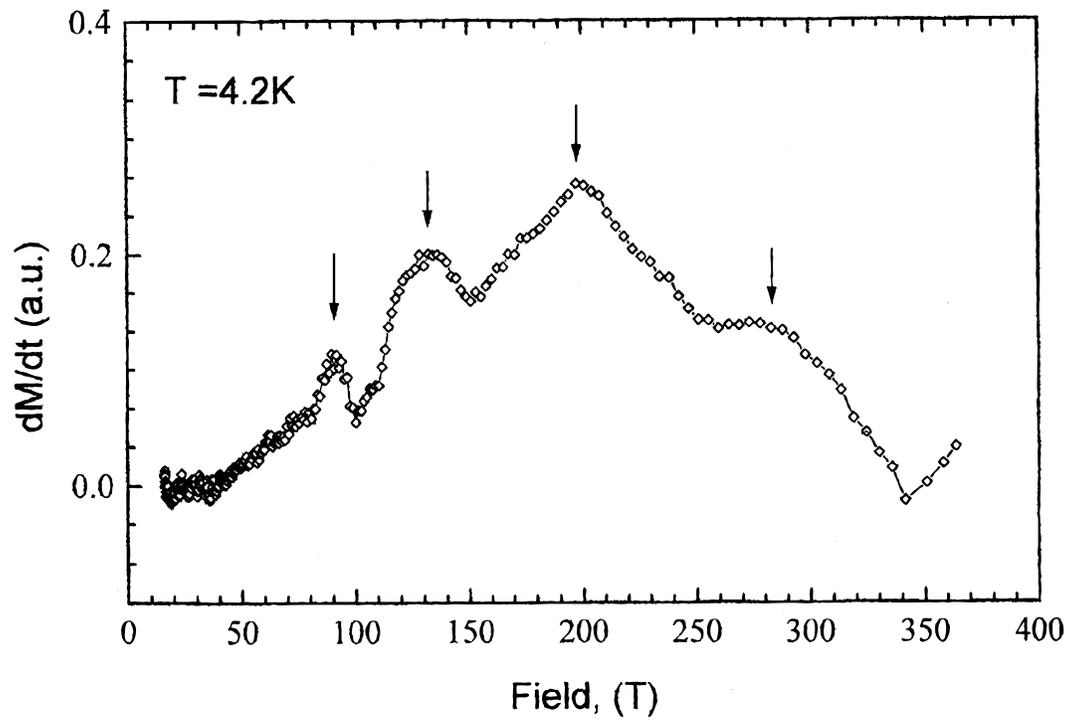}
\caption{Experimental dependence $\chi\left( H \right)$ of
magnetic susceptibility on magnetic field.}
\end{figure*}

\end{document}